\documentclass[preprint,cha,onecolumn,superscriptaddress,preprintnumbers,showpacs,amsmath,amssymb]{revtex4-1}
\usepackage{bm}
\usepackage[colorlinks=true,linkcolor=blue,citecolor=blue]{hyperref}
\usepackage{amsthm}
\usepackage{amsfonts}
\usepackage{enumerate}
\usepackage{latexsym}
\usepackage{float}
\usepackage{ifpdf}
\usepackage{graphicx}
\usepackage{makeidx}
\expandafter\ifx\csname package@font\endcsname\relax\else
 \expandafter\expandafter
 \expandafter\usepackage
 \expandafter\expandafter
 \expandafter{\csname package@font\endcsname}
 \fi
\usepackage{color}
\usepackage{times}
\usepackage{multirow}
\usepackage{physics}

\begin{document}

\title{Dirac nodal lines in the quasi-one-dimensional ternary telluride TaPtTe$_5$}
\author{Shaozhu Xiao}
\email{xiaoshaozhu@nimte.ac.cn}
\affiliation{Ningbo Institute of Materials Technology and Engineering, Chinese Academy of Sciences, Ningbo 315201, China}
\author{Wen-He Jiao}
\email{whjiao@zjut.edu.cn}
\affiliation{Key Laboratory of Quantum Precision Measurement of Zhejiang Province, Department of Applied Physics, Zhejiang University of Technology, Hangzhou 310023, China}
\author{Yu Lin}
\affiliation{Ningbo Institute of Materials Technology and Engineering, Chinese Academy of Sciences, Ningbo 315201, China}
\affiliation{Fujian Provincial Collaborative Innovation Center for Advanced High-Field Superconducting Materials and Engineering, Fuzhou 350117, China}
\affiliation{College of Physics and Energy, Fujian Normal University, Fuzhou 350117, China}
\author{Qi Jiang}
\affiliation{State Key Laboratory of Functional Materials for Informatics, Shanghai Institute of Microsystem and Information Technology, Chinese Academy of Science, Shanghai 200050, China}
\author{Xiufu Yang}
\affiliation{Ningbo Institute of Materials Technology and Engineering, Chinese Academy of Sciences, Ningbo 315201, China}
\author{Yunpeng He}
\affiliation{Ningbo Institute of Materials Technology and Engineering, Chinese Academy of Sciences, Ningbo 315201, China}
\author{Zhicheng Jiang}
\affiliation{State Key Laboratory of Functional Materials for Informatics, Shanghai Institute of Microsystem and Information Technology, Chinese Academy of Science, Shanghai 200050, China}
\author{Yichen Yang}
\affiliation{State Key Laboratory of Functional Materials for Informatics, Shanghai Institute of Microsystem and Information Technology, Chinese Academy of Science, Shanghai 200050, China}
\author{Zhengtai Liu}
\affiliation{State Key Laboratory of Functional Materials for Informatics, Shanghai Institute of Microsystem and Information Technology, Chinese Academy of Science, Shanghai 200050, China}
\author{Mao Ye}
\affiliation{State Key Laboratory of Functional Materials for Informatics, Shanghai Institute of Microsystem and Information Technology, Chinese Academy of Science, Shanghai 200050, China}
\author{Dawei Shen}
\affiliation{State Key Laboratory of Functional Materials for Informatics, Shanghai Institute of Microsystem and Information Technology, Chinese Academy of Science, Shanghai 200050, China}
\author{Shaolong He}
\email{shaolonghe@nimte.ac.cn}
\affiliation{Ningbo Institute of Materials Technology and Engineering, Chinese Academy of Sciences, Ningbo 315201, China}
\affiliation{University of Chinese Academy of Sciences, Beijing 100049, China}


\begin{abstract}

A Dirac nodal-line phase, as a quantum state of topological materials, usually occur in three-dimensional or at least two-dimensional materials with sufficient symmetry operations that could protect the Dirac band crossings. Here, we report a combined theoretical and experimental study on the electronic structure of the quasi-one-dimensional ternary telluride TaPtTe$_5$, which is corroborated as being in a robust nodal-line phase with fourfold degeneracy. Our angle-resolved photoemission spectroscopy measurements show that two pairs of linearly dispersive Dirac-like bands exist in a very large energy window, which extend from a binding energy of $\sim$ 0.75 eV to across the Fermi level. The crossing points are at the boundary of Brillouin zone and form Dirac-like nodal lines. Using first-principles calculations, we demonstrate the existing of nodal surfaces on the $k_y = \pm \pi$ plane in the absence of spin-orbit coupling (SOC), which are protected by nonsymmorphic symmetry in TaPtTe$_5$. When SOC is included, the nodal surfaces are broken into several nodal lines. By theoretical analysis, we conclude that the nodal lines along $Y$-$T$ and the ones connecting the $R$ points are non-trivial and protected by nonsymmorphic symmetry against SOC.

\end{abstract}

\maketitle
\newpage

\section{INTRODUCTION}

Since the remarkable discovery of topological insulators, searching for new types of topological materials and studying of topologically non-trivial states of matters have been a hot topic in the field of condensed-matter physics. During the past decade, the experimentally confirmed topological materials have been extended from insulators \cite{Science-YLC-2009,Science-YLC-2010} to semimetals \cite{PRB-2011-AAB,PRL-SMY-2012,PRB-ZW-2012,Science-ZKL-2014,Science-SYX-2015-Dirac,PRB-XW-2011,NC-SMH-2015,PRX-WH-2015,Science-SYX-2015-Weyl}, and even superconductors\cite{Science-DW-2018,Science-PZ-2018}. In contrast to topological insulators that host robust metallic surface states along with an insulating bulk, topological semimetals (TSMs) host metallic surface states and a semi-metallic bulk which exhibits a small or vanishing density of states near the Fermi level, originating from the topology- or symmetry- protected band crossings near the Fermi level. After nearly a decade of rapid development, the number of experimental realized TSMs has been numerous. They can be categorized on the basis of the degeneracy or the dimension of the band crossings in the momentum space, with the former leading to Dirac semimetals\cite{PRL-SMY-2012,PRB-ZW-2012,Science-ZKL-2014,Science-SYX-2015-Dirac}, Weyl semimetals\cite{PRB-XW-2011,NC-SMH-2015,PRX-WH-2015,Science-SYX-2015-Weyl} and triple-point semimetals\cite{PRB-HW-2016,PRX-ZZ-2016,NC-WG-2018} and the latter leading to zero-dimensional (0D) nodal point semimetals, one-dimensional (1D) nodal line semimetals\cite{NC-2016-GB,PRB-2016-MN,NC-2016-LMS,NP-2018-SP,SA-2019-BB,PRL-2020-YKS,NM-2020-TYY} and two-dimensional (2D) nodal surface semimetals\cite{PRB-2016-QFL,PRB-2018-WW}.

Crystallographic symmetries play an important role in TSMs, in terms of protecting the band crossings against spin-orbit coupling (SOC). For instance, nonsymmorphic symmetries give opportunities to realize robust band crossings at the Brillouin zone boundary\cite{PRB-2016-YXZ,PRB-2017-BJY,APX-2018-SYY}, i.e. symmetry-enforced band crossings, which can generate nodal lines or nodal surfaces even in low-dimensional or strong anisotropic systems. $MM'$Te$_5$ ($M$=Nb, Ta; $M'$=Ni, Pd, Pt) is such a type of quasi-one-dimensional ternary telluride that host nonsymmorphic symmetries\cite{EJIC-1999-PA,PRB-2020-WHJ,JPCL-2020-CX,PRB-2021-WHJ,PRB-2021-ZC,PRB-2021-ZH}. The experimental characterizations and first-principles calculations indicate that Ta$M'$Te$_5$ ($M'$=Ni, Pd, Pt) all possess nontrivial band topology\cite{PRB-2020-WHJ,JPCL-2020-CX,PRB-2021-WHJ,PRB-2021-ZC}. According to the symmetry analysis of their crystal structures, $MM'$Te$_5$ is supposed to have nodal surfaces at the Brillouin zone boundary without considering SOC. If taking SOC into account, the nodal surfaces would degenerate into several nodal lines. To date, the direct experimental studies of the electronic structures of $MM'$Te$_5$ are still rare except a recent one on TaNiTe$_5$ \cite{PRB-2021-ZH} by angle-resolved photoemission spectroscopy (ARPES).

Here, we report a comprehensive study on the electronic structure of TaPtTe$_5$ by means of ARPES measurements and first-principles calculations. We observed Dirac-like nodal lines at the Brillouin zone boundary of TaPtTe$_5$, which have linear dispersions extending from a binding energy of $\sim$ 0.75 eV to across the Fermi level. In combination with theoretical calculations, we get into a conclusion that the observed Dirac-like nodal lines by ARPES are the bulk band features at the Brillouin zone boundary, which are gapless without considering SOC, and a small gap is opened at these Dirac points when taking SOC into account. By theoretical analysis, we also find that the nodal lines along $Y$-$T$ path and the nodal lines connecting the $R$ points are protected by nonsymmorphic symmetry and are robust against SOC.

\section{EXPERIMENTS AND METHODS}
Single crystals of TaPtTe$_5$ were grown using a Te self-flux method\cite{PRB-2021-WHJ}. Powders of the elements Ta (99.97\%), Pt (99.98\%) and Te (99.99\%) were mixed in a molar ratio of Ta : Pt : Te = 1 : 1 :10. The mixture was loaded into a quartz ampoule in a highly-purified argon atmosphere glove box, and then evacuated and sealed. The quartz ampoule was firstly heated up to 773 K and held for 24 h, and secondly heated to 1073 K and held for another 24 h, and then quickly heated to 1273 K and held for 4 days,  and finally slowly cooled down to room temperature in 24 h.

ARPES measurements were carried out on beamline BL03U at the Shanghai Synchrotron Radiation Facility (SSRF) with a Scienta DA30 analyzer\cite{SSRF-1,SSRF-2}. The energy resolution was better than 5 meV and the angular resolution was better than 0.1$^\circ$\cite{SSRF-2}. The base pressure was better than $8 \times 10^{-11 }$~mbar. All the samples were cleaved and measured at a temperature of about 20~K.

The theoretical electronic structures of TaPtTe$_5$ were obtained by employing relativistic first-principle calculations based on the density-functional theory as implemented in the \textsc{Quantum ESPRESSO} code\cite{JPCM-QE-2009}. The core electrons were described by the projector augmented wave method (PAW)\cite{PRB-PAW-1994}, and the exchange correlation energy was approximated by the Generalized Gradient Approximation (GGA) using the PBE functional\cite{PRL-PBE-1996}. The plane-wave kinetic cut-off energy was set to be 60 Ry, and the Brillouin zone was sampled with a \textbf{k} mesh of $8 \times 8 \times 4$. 
In order to obtain surface spectrum, a first-principles tight-binding model Hamilton based on maximally localized Wannier functions was constructed by fitting the DFT band structures using \textsc{Wannier90} code\cite{JPCM-wannier90-2020}. The Ta-\textit{d} orbital, Pt-\textit{d} orbital and Te-\textit{p} orbital were used for the initial projection. The surface spectra of TaPtTe$_5$ were calculated using the Green function method as implemented in \textsc{WannierTools} package\cite{CPC-wt-2018}.

\section{RESULTS AND DISCUSSION}

\begin{figure*}[h]
\includegraphics[width=0.95\textwidth]{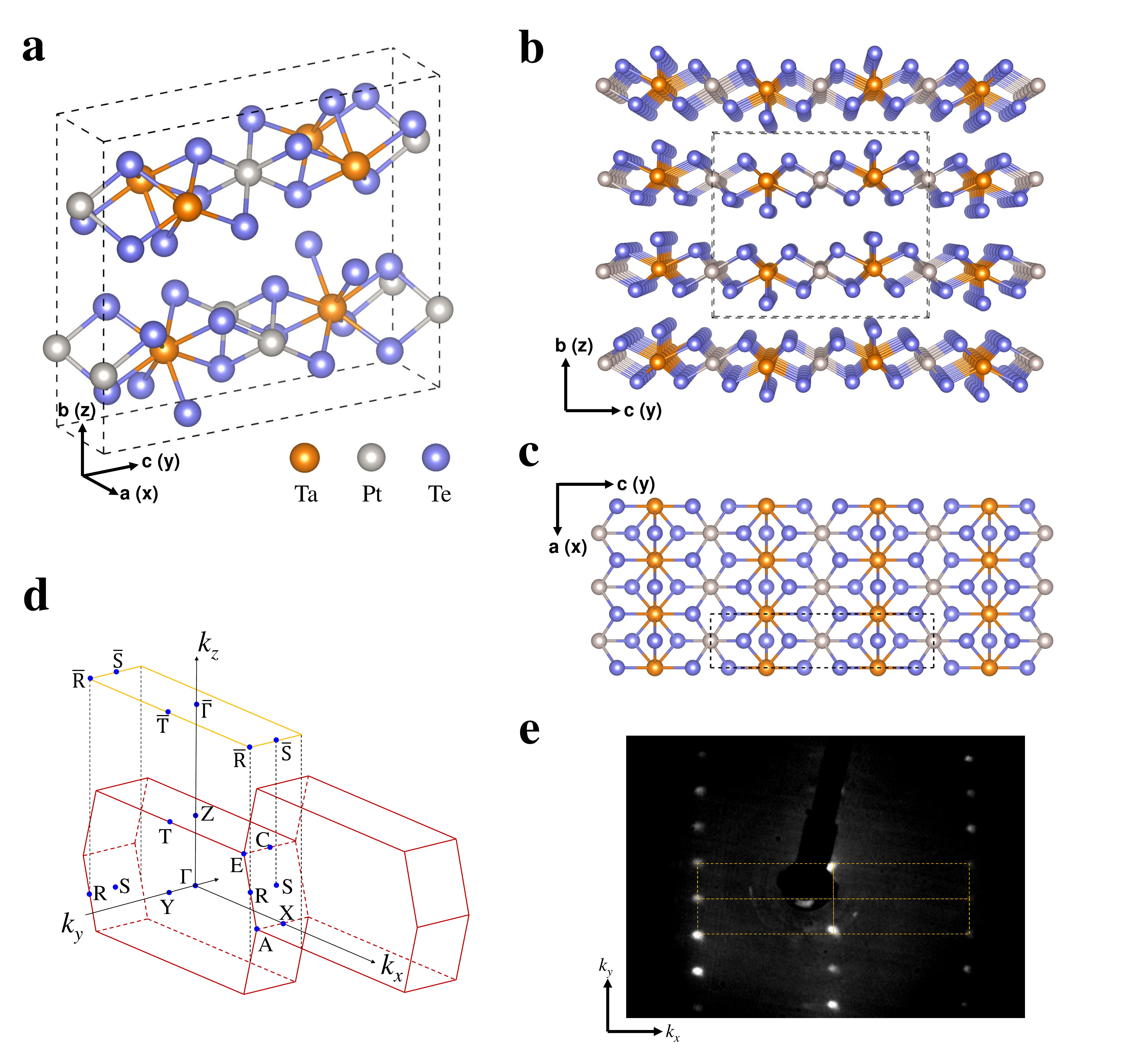}
\caption{\label{Fig1}Crystal structure and sample characterization of TaPtTe$_5$. (a) Conventional unit cell of TaPtTe$_5$. For convenience, the $a$, $c$ and $b$ axes are specified as $x$, $y$ and $z$ axes, respectively. (b) Perspective view of the crystal structure of TaPtTe$_5$ along the $x$ axis. (c) View of a TaPtTe$_5$ layer along the $z$ axis. (d)Three dimensional bulk Brillouin zone (red lines) and the projected (001) surface Brillouin zone (yellow lines) with the high-symmetry points labelled. (e) Low energy electron diffraction (LEED) pattern of the cleaved (001) surface.}
\end{figure*}

\noindent\textbf{A. Crystal structure and symmetry of TaPtTe$_5$}

TaPtTe$_5$ crystallizes in an orthorhombic layered structure with space group $Cmcm$ (No. 63), and lattice constants are $a = 3.729$~\AA, $b = 13.231$~\AA~and $c = 15.452$~\AA [Fig.~\ref{Fig1}(a)].  For convenience, we specify the $a$, $c$ and $b$ axes as $x$, $y$ and $z$ axes, respectively. As seen in Figs.~\ref{Fig1}(a-c), the unit cell of TaPtTe$_5$ consists of two TaPtTe$_5$ layers, and each layer can be described as a series of bicapped trigonal prismatic TaTe$_5$ chains glued by Pt atoms. The TaPtTe$_5$ layers are stacked by shifting along the chain direction ($x$ direction) by $\frac{a}{2}$  with respect to the adjacent layer. The layers are bounded together via van der Waals-type weak interaction and are thus easily cleaved. The natural cleaved surface is (001) surface. Figure~\ref{Fig1}(d) plots the bulk Brillouin zone of the primitive cell and the projected (001) surface Brillouin zone. Figure~\ref{Fig1}(e) shows the low energy electron diffraction (LEED) pattern of the cleaved surface of TaPtTe$_5$. The rectangular arrangement of LEED spots confirms the cleaved (001) surface, and the long side of the rectangle corresponds to the chain direction ($x$ direction) while the short side corresponds to the interchain direction ($y$ direction).

For the sake of later discussion, we now look into the symmetry of TaPtTe$_5$. As seen in Figs.~\ref{Fig1}(a-c), it is obvious that TaPtTe$_5$ has inversion symmetry $\mathcal{P}$ with the inversion center sitting on a Pt atom or the midpoint of the bond of a nearest pair of Pt atoms. TaPtTe$_5$ contains no magnetic elements, and naturally possesses time-reversal symmetry $\mathcal{T}$. TaPtTe$_5$ is also invariant under two nonsymmorphic operations. One is glide mirror $\widetilde{\mathcal{M}}_{z}$ (Here the tilde above a symbol indicates a nonsymmorphic symmetry), and it operates as a mirror reflection (the mirror plane is perpendicular to the $z$ axis and goes through the center of Ta atom) followed by a half of lattice translation along the $y$ axis. The other is glide mirror $\widetilde{\mathcal{M}}_{y}$, and it operates as a mirror reflection (the mirror plane is perpendicular to the $y$ axis and goes through the center of Pt atom or Ta atom) followed by a half of lattice translation along the $y$ axis.

\begin{figure*}[p]
\includegraphics[width=0.9\textwidth]{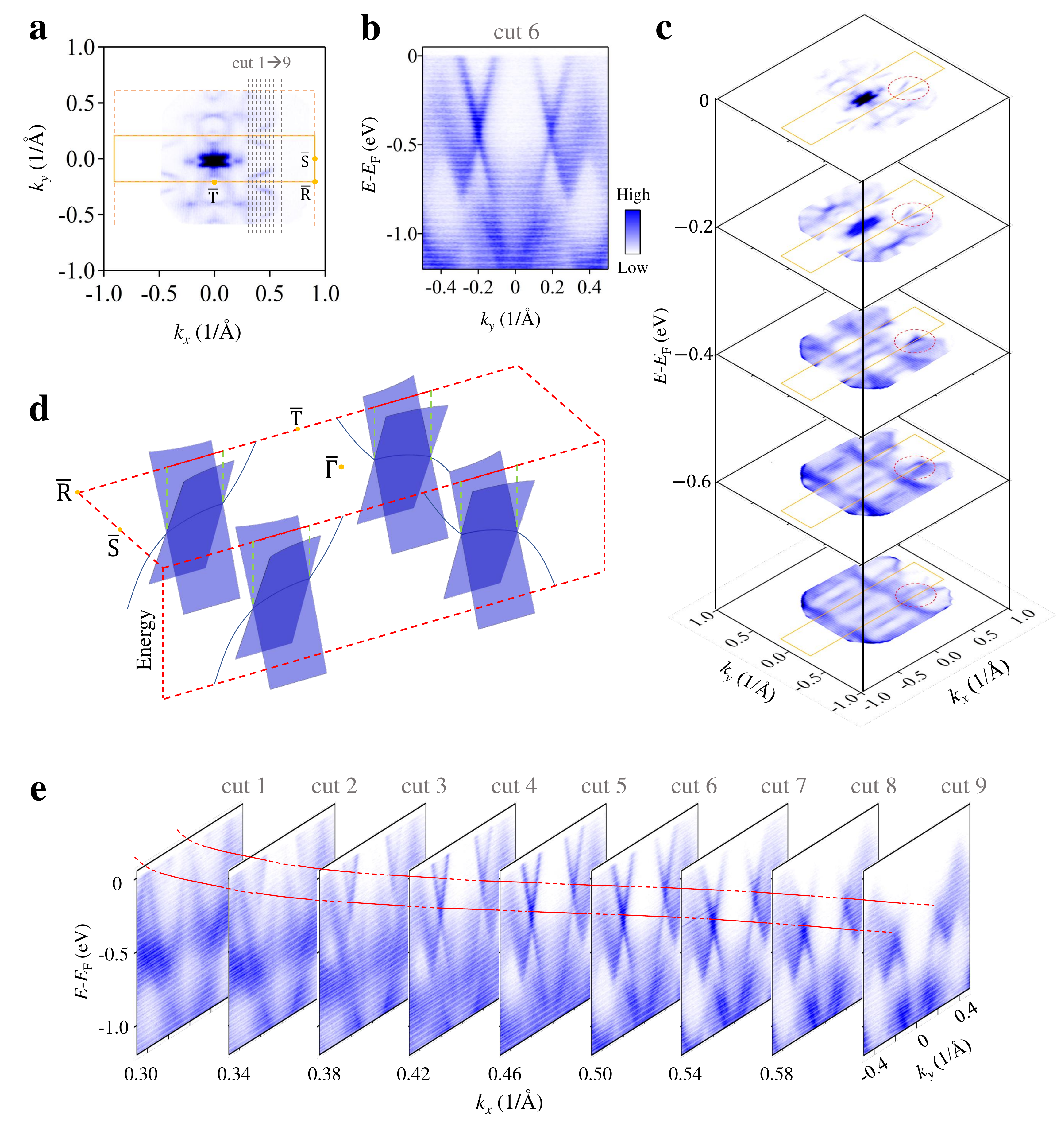}
\caption{\label{Fig2}Electronic structures of TaPtTe$_5$ measured by ARPES ($h\nu = 28.5$~eV). (a) ARPES Fermi surface of TaPtTe$_5$. The yellow lines represent the surface Brillouin zone. The grey dashed lines in (a) indicate the momentum cuts along which the band dispersions in (b) and (e) are taken. (b)  A representative Dirac-like band dispersion along the ``cut 6'' grey line as indicated in (a). (c) Stacking plot of constant-energy contours at different binding energies. (d) Schematic plot of the measured Dirac-like nodal lines in the surface Brillouin zone. (e) A series of band dispersions along the grey lines as indicated in (a) to reveal the nodal lines.}
\end{figure*}

\noindent\textbf{B. Dirac nodal line of TaPtTe$_5$ probed by ARPES}

In order to study the electronic structures of TaPtTe$_5$, we cleaved the samples \textit{in-situ} and performed ARPES measurements on the cleaved surfaces, and the results are shown in Fig.~\ref{Fig2}. Figure \ref{Fig2}(a) shows the ARPES measured Fermi surface and Fig. \ref{Fig2}(c) shows the stacking plot of constant-energy contours at different binding energies to exhibit the evolution of band structures. In the vicinity of the Fermi level, the spectral weight locates majorly around the $\overline{\Gamma}$ point, and spreads along the short side  (corresponding to the direction of inter-chain) of the rectangle-shaped surface Brillouin zone. At the boundary of Brillouin zone (line $\overline{T}$-$\overline{R}$), two curves get close, merge into one curve at $\sim$ 0.4 eV and then separate again with increasing the binding energy, as indicated by the red dashed circles in Fig.~\ref{Fig2}(c), which is a signature of the possible presence of nodal line. 

Figure \ref{Fig2}(b) shows a representative band dispersion image of the nodal line, which is along the ``cut 6'' grey line (corresponding to $k_x$ = 0.5 \AA$^{-1}$) as indicated in Fig.~\ref{Fig2}(a). Two pairs of linearly dispersive Dirac-like bands  exist in a very large energy window. The linear dispersions extend from a binding energy of $\sim$ 0.75 eV to across the Fermi level, and these bands have a Fermi velocity of about $0.6\times 10^6$ m/s which is comparable to that in graphene ($\sim 1\times 10^6$ m/s). Each pair of Dirac-like bands cross each other at a binding energy of about 0.4 eV. The crossing points are at the boundary of Brillouin zone (on line $\overline{T}$-$\overline{R}$) , and give rise to the observed nodal lines. 

To reveal the presence of these nodal lines as a whole, we plot in Fig.~\ref{Fig2}(e) the band dispersions along a series of parallel cuts, as indicated by the grey dashed lines in Fig.~\ref{Fig2}(a). Between $k_x$ = 0.42 \AA$^{-1}$ (cut 4) and $k_x$ = 0.58 \AA$^{-1}$ (cut 8), the crossing points of the Dirac-like bands sit almost around 0.4 eV and the nodal lines have little dispersion. When going from $k_x$ = 0.42 \AA$^{-1}$ to $k_x$ = 0, the crossing points gradually move upward and cross the Fermi level at around $k_x$ = 0.3 \AA$^{-1}$ (cut 1). In contrast, when going from $k_x$ = 0.58 \AA$^{-1}$ to the $\overline{R}$ point, the crossing points sharply move toward high binding energy. A schematic plot of the measured Dirac-like nodal lines in the surface Brillouin zone is shown in Fig. \ref{Fig2}(d).

\noindent\textbf{C. Calculation in the absence of SOC}

\begin{figure*}[]
\includegraphics[width=\textwidth]{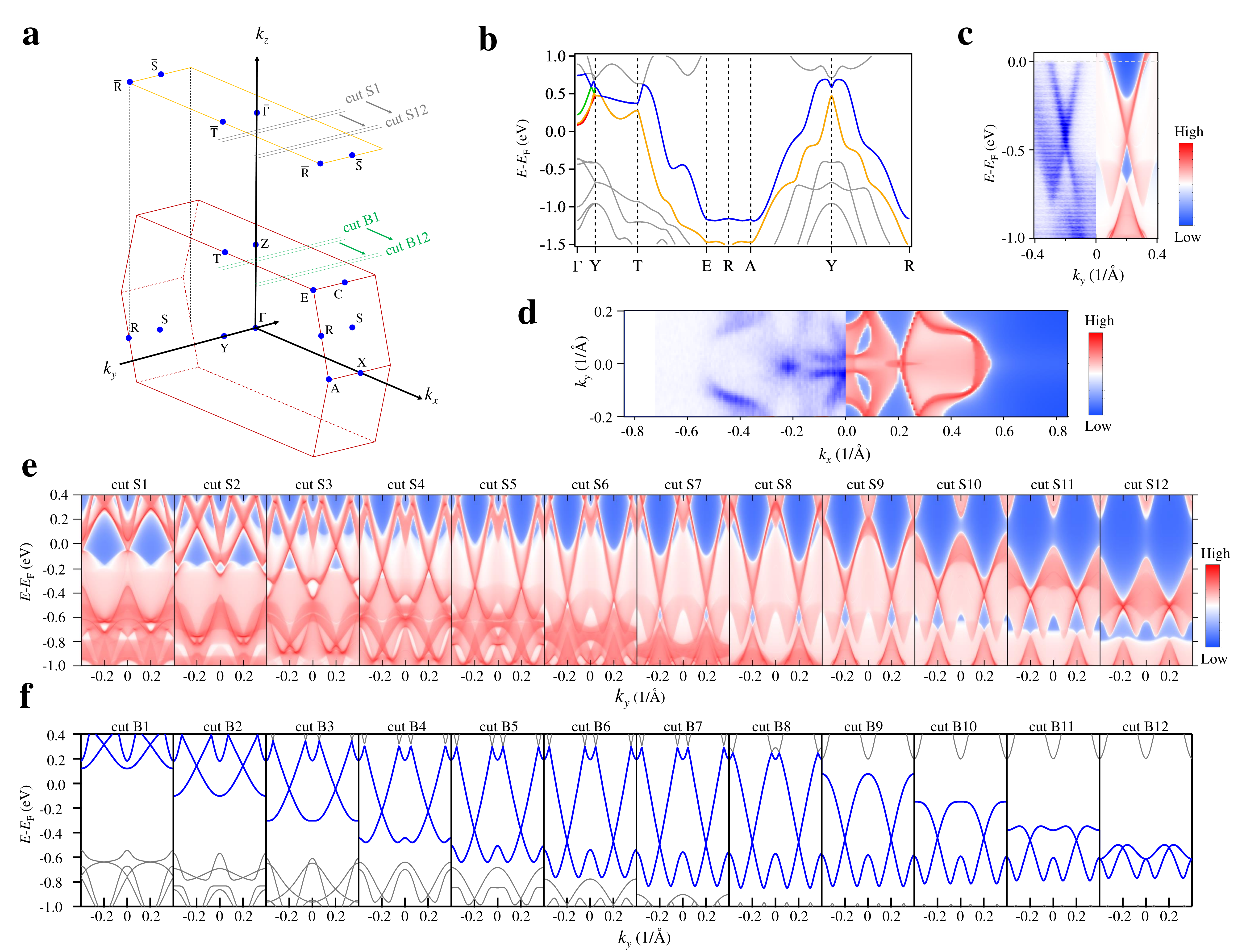}
\caption{\label{Fig3}Electronic structures of TaPtTe$_5$ obtained by theoretical calculation without considering the SOC. (a) Bulk Brillouin zone (red lines) and surface Brillouin zone (yellow lines). (b) Calculated bands along selected high-symmetry lines in the absence of SOC. (c) ARPES measured band dispersions (left side) and the corresponding calculated results (right side), with $k_x$ = $\sim$0.5 \AA$^{-1}$.  (d) ARPES measured (left side) and calculated (right side) Fermi surface. (e) Calculated surface spectra along the lines in surface Brillouin zone as indicated by the grey lines in (a), which include the contributions from both the projections of bulk bands and the surface states. (f) Calculated bulk bands along the lines on the $k_z = \pi$ plane of bulk Brillouin zone as indicated by the green lines in (a).}
\end{figure*}

To study the origin of the ARPES observed Dirac-like nodal lines, we theoretically calculated the band structures of TaPtTe$_5$. We first consider the case in the absence of SOC. In such case, we can find that all bands are two-fold degenerate without considering the spin (four-fold degenerate when considering the spin) on the $k_y = \pm \pi$ plane [$Y$-$T$-$E$-$R$-$A$ plane as shown in Fig.~\ref{Fig3}(a)], based on symmetry analysis as presented in the supplementary material\cite{supplementary}. That is, in the absence of SOC, the bands on the $k_y = \pm \pi$ plane form nodal surfaces, which are protected by nonsymmorphic symmetry $\widetilde{\mathcal{M}}_{z}$ and $\widetilde{\mathcal{M}}_{y}$.

Figure \ref{Fig3}(b) shows the calculated bulk bands of TaPtTe$_5$ along some high-symmetry lines in the absence of SOC. Along the $\Gamma$-$Y$, which is outside of the $k_y = \pm \pi$ plane, the bands are not degenerate, while approaching the $Y$ point on the $k_y = \pm \pi$ plane, the bands become degenerate. All bands along the lines on the $k_y = \pm \pi$ plane are two-fold degenerate, which is consistent with the conclusion from symmetry analysis\cite{supplementary} that  nodal surfaces exist on the $k_y = \pm \pi$ plane. Figure \ref{Fig3}(c) shows a comparison between the ARPES measured band dispersions (left side) and the corresponding calculated result [right side, along cut S9 ($k_x = 0.6~\overline{T}\overline{R} \sim$ 0.5 \AA$^{-1}$) as indicated in Fig. \ref{Fig3}(a)], the details of which are almost the same. The good agreement is also found between the ARPES measured Fermi surface and the calculated one, as shown in Fig.~\ref{Fig3}(d). 
Figure~\ref{Fig3}(e) displays the calculated surface spectra along several selected lines in the surface Brillouin zone [from cut S1 to S12 as indicated in Fig.~\ref{Fig3}(a), corresponding to $k_x = 0.24~\overline{T}\overline{R}$, $k_x = 0.28~\overline{T}\overline{R}$, ..., and $k_x = 0.68~\overline{T}\overline{R}$, respectively], which are basically in line with our experimental results. These calculated surface spectra include the contributions from both the projections of bulk bands and the surface states. By comparison with the bulk bands on the $k_z = \pm \pi$ plane [Fig.~\ref{Fig3}(f)] and other planes [see Fig. S3 in Ref. \onlinecite{supplementary}], we can find that the surface spectral weight is dominated by the projections of the bulk bands at the boundary of Brillouin zone ($T$-$E$-$C$-$Z$ plane, $k_z = \pm \pi$ plane). The ARPES observed Dirac-like nodal lines are a part of the nodal surfaces, and they are mainly from the boundary of Brillouin zone (on line $T$-$E$, the intersection of the $k_y = \pm \pi$ and $k_z = \pm \pi$ planes).

We also conducted the photon-energy-dependent ARPES measurements on TaPtTe$_5$ to reveal the $k_z$ dependence of band structures, and the results are shown in the supplementary Figs. S1 and S2\cite{supplementary}. However, no significant changes of the Fermi surface and the constant-energy contours are identified. The Dirac-like bands are always observed with photon energy increased from 22 eV to 50 eV, except the sharpness gradually weakens with the increase of photon energy after 30 eV. 
The ARPES measured Fermi surfaces (first row of Fig. S1 in Ref.  \onlinecite{supplementary}) with different photon energies are almost the same and agree well with the calculated one [right side of Fig.~\ref{Fig3}(d)]. The calculated Fermi surface comprises of the surface states and the whole projections of bulk states (the calculated three-dimensional Fermi surfaces in the bulk Brillouin zone can be seen in Fig. S5 in Ref. \onlinecite{supplementary}). Therefore, we can conclude that ARPES measurments with a fixed photon energy in range of 20 $\sim$ 50 eV cover the band structures across the entire $k_z$ range due to the $k_z$ broadening effect in TaPtTe$_5$. The $k_z$ broadening effect is common in low-photon-energy ARPES measurements\cite{PRB-2021-Xiao,PRB-2019-Wu,PRB-2018-Wang,PRB-2016-Niu,PRM-2018-Jiang,PRB-2017-Oinuma} and usually leads to no significant changes of band structures in the photon-energy-dependent ARPES measurements.

\noindent\textbf{D. Calculation in the presence of SOC}

\begin{figure*}[]
\includegraphics[width=0.8\textwidth]{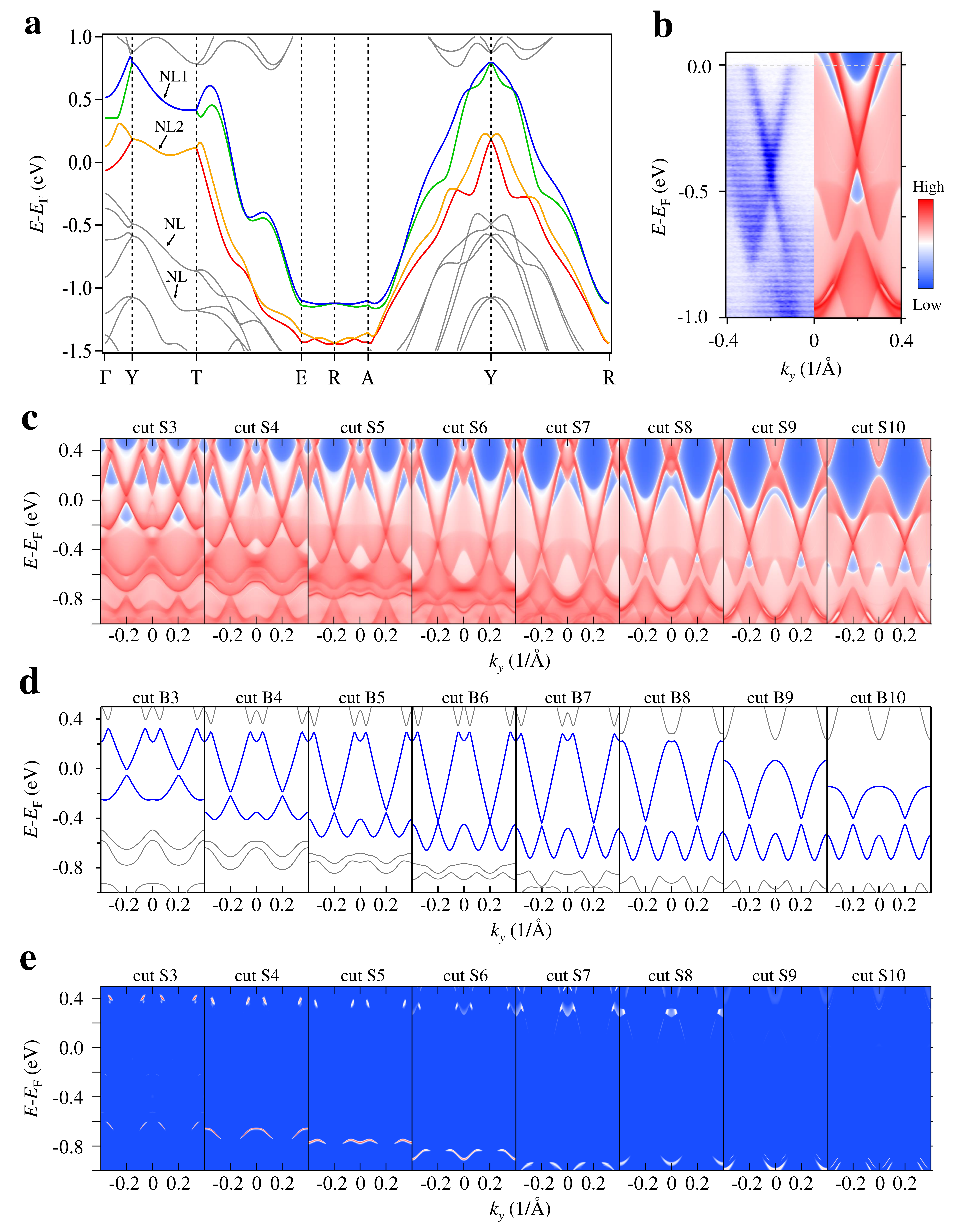}
\caption{\label{Fig4}Electronic structures of TaPtTe$_5$ obtained by theoretical calculation when considering SOC. (a) Calculated bands along selected high-symmetry lines in the presence of SOC. The bands (along $Y$-$T$) indicated by black arrows are Dirac nodal lines protected by symmetry against SOC. (b) ARPES measured band dispersions (left side) and the corresponding calculated result (right side), with $k_x$ = $\sim$0.5 \AA$^{-1}$. (c) Calculated surface spectra along the lines in surface Brillouin zone (cut S3 to S10) as indicated by the grey lines in Fig.~\ref{Fig3}(a). (d) Calculated bulk bands along the lines on the $k_z = \pi$ plane of bulk Brillouin zone as indicated by the green lines in Fig.~\ref{Fig3}(a). (e) The contributions of surface states extracted from Fig.~\ref{Fig4}(c).}
\end{figure*}

When SOC is included, the nodal surfaces on the $k_y = \pm \pi$ plane are broken. Figure \ref{Fig4}(a) shows the calculated bulk bands of TaPtTe$_5$ in the presence of SOC. In comparison with those in Fig.~\ref{Fig3}(b) in the absence of SOC, we can find that the bands along some lines on the $k_y = \pm \pi$ plane are split, suggesting the breaking of the nodal surfaces on the $k_y = \pm \pi$ plane. Figure~\ref{Fig4}(b) shows a comparison between ARPES measured and calculated band dispersions (cut S9 with $k_x$ = $\sim$0.5 \AA$^{-1}$) for the case with SOC. Compared with the non-SOC case, the calculated results do not significantly change and also match the experiments satisfactorily. Figure \ref{Fig4}(c) shows the calculated surface spectra along several selected lines [cut S3 to S10 as indicated in Fig.~\ref{Fig3}(a)] in the surface Brillouin zone for the case with SOC. Similar to those in Fig.~\ref{Fig3}(e), the most prominent features in Fig.~\ref{Fig4}(c) are also the Dirac-like bands. The major difference is that in the latter some Dirac points are gapped. As discussed in the non-SOC case, the surface spectral weight is dominated by the projections of the bulk bands on the $k_z = \pm \pi$ plane, and Fig.~\ref{Fig4}(d) shows the bulk bands along the corresponding lines on the $k_z = \pm \pi$ plane (the bulk bands on other planes are shown in Fig. S4 in Ref. \onlinecite{supplementary}). The introduction of SOC makes some Dirac points be gapped, and the gaps on the $k_z = \pm \pi$ plane are relatively small in comparison with those on other planes. For instance, the gaps in cut B6 in Fig.~\ref{Fig4}(d) are as small as 1.8 meV. Figure~\ref{Fig4}(e) shows the contributions of the surface states extracted from the calculated surface spectra in Fig.~\ref{Fig4}(c), which can also rule out the possibility that the ARPES observed Dirac-like band structures are derived from the surface states.

\begin{figure*}[h]
\includegraphics[width=\textwidth]{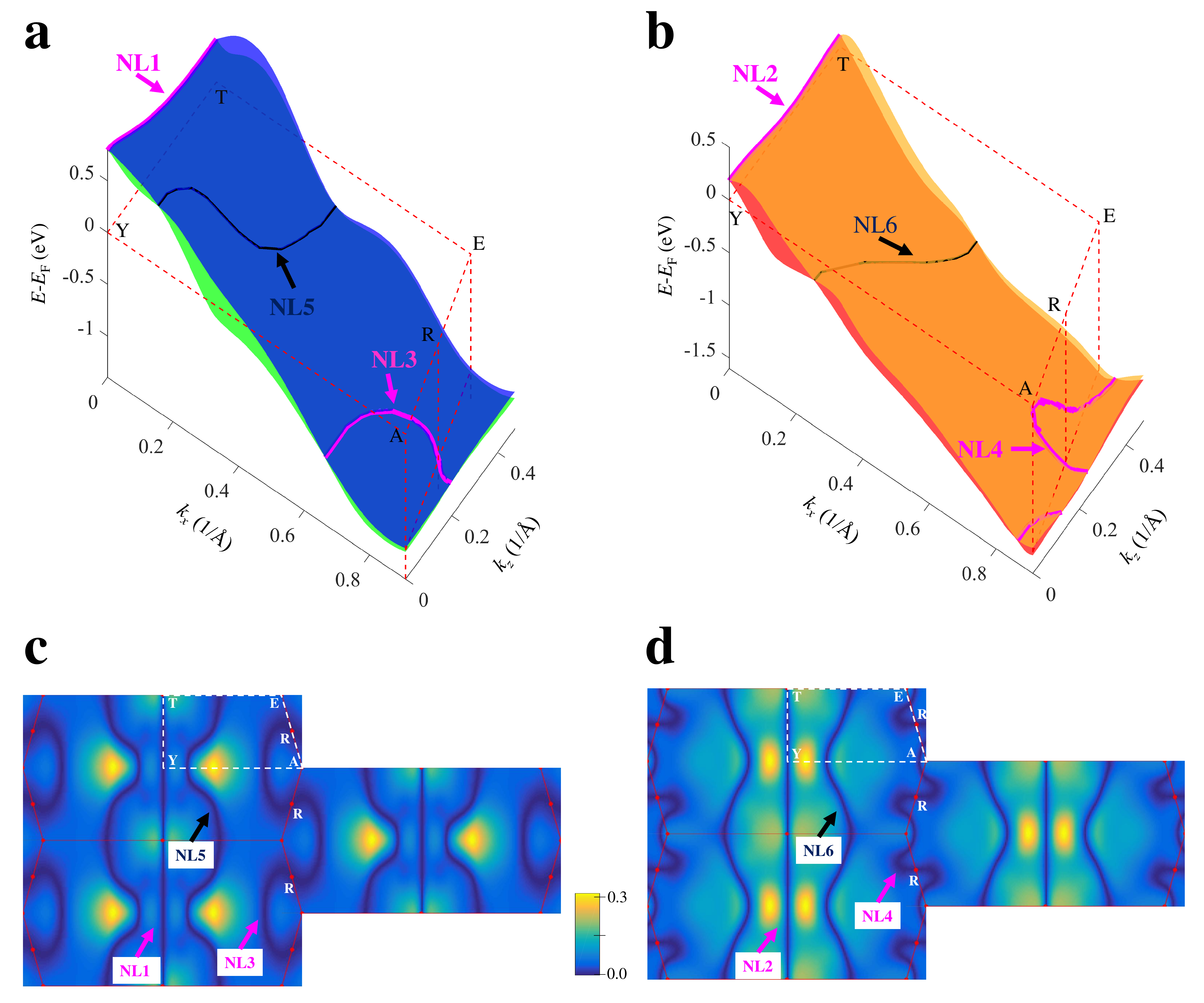}
\caption{\label{Fig5} Nodal lines on the $k_y = \pm \pi$ plane. (a, b) The upmost two pairs of bands that cross the Fermi level, and (c, d) the difference between each pair. six nodal lines on the $k_y = \pm \pi$ plane are shown, four of which as shown by pink lines in (a, b) and as indicated by pink arrows in (c, d) are protected by nonsymmorphic symmetry.}
\end{figure*}

Although the nodal surfaces on the $k_y = \pm \pi$ plane are broken by SOC,  some nodal lines are retained. Figures \ref{Fig5}(a) and \ref{Fig5}(b) shows the upmost two pairs of bands that cross the Fermi level on the $k_y = \pm \pi$ plane, and the difference of them are shown in Figs.~\ref{Fig5}(c) and \ref{Fig5}(d), respectively. As seen in Fig. \ref{Fig5}, each pair of these bands form three nodal lines. 
Since the nonsymmorphic operation $\widetilde{\mathcal{M}}_{y}$ keeps the system invariant and commutes with the Hamiltonian, the Bloch state can be chosen to be their common eigenstate $\ket{u}$. Thus, on $Y$-$T$ path, we can find two energetically degenerate Kramers pairs $\ket{u}$, $\mathcal{PT} \ket{u}$ and $\widetilde{\mathcal{M}}_{z}\mathcal{T}\ket{u}$, $\widetilde{\mathcal{M}}_{z}\mathcal{P}\ket{u}$ with opposite $\widetilde{\mathcal{M}}_{y}$ eigenvalues (see the supplementary material\cite{supplementary} for details). That is, all bands along $Y$-$T$ path are four-fold degenerate and form nodal lines, of which the four-fold degeneracy is essential and is symmetry protected. Similarly, for the time-reversal invariant momentum (TRIM) points $R$, the two energetically degenerate Kramers pairs with opposite $\widetilde{\mathcal{M}}_{y}$ eigenvalues can be found as $\ket{u}$, $\mathcal{PT} \ket{u}$ and $\mathcal{T}\ket{u}$, $\mathcal{P}\ket{u}$\cite{supplementary}. So the four-fold degeneracy of bands at $R$ points is also symmetry protected, and the nodal lines connecting $R$ points are movable but not removable. Therefore, for the nodal lines shown in Fig.~\ref{Fig5}, the ones along $Y$-$T$ path (NL1 and NL2) and the ones connecting $R$ points (NL3 and NL4) are non-trivial and are protected by nonsymmorphic symmetry. In TaPtTe$_5$, the nodal lines connecting $R$ points are all far away from the Fermi level. However, along $Y$-$T$ path, there are several nodal lines in the vicinity of the Fermi level, e.g., NL1 and NL2 shown in Fig.~\ref{Fig5} and Fig.~\ref{Fig4}(a). These nodal lines may contribute to the transport properties of TaPtTe$_5$.

\section{CONCLUSION}
In summary, the electronic structure of TaPtTe$_5$ were comprehensively investigated by means of ARPES in combination with theoretical analysis and DFT calculations. By using ARPES, we observed Dirac-like nodal lines at the boundary of surface Brillouin zone of TaPtTe$_5$. The Dirac-like bands exist in a very large energy window, which extend from a binding energy of $\sim$ 0.75 eV to across the Fermi level. Based on theoretical analysis, we conclude that there are nodal surfaces existing on the $k_y = \pm \pi$ plane in the absence of SOC. The ARPES observed Dirac-like nodal lines are a part of the nodal surfaces, and they are the bulk band features from the Brillouin zone boundary. When SOC is included, the nodal surfaces on the $k_y = \pm \pi$ plane are broken, and small gaps are opened at the crossing points of the ARPES observed Dirac-like nodal lines. Even though the nodal surfaces on the $k_y = \pm \pi$ plane are broken in the presence of SOC, there are still several non-trivial nodal lines. The nodal lines along $Y$-$T$ path and the ones connecting the $R$ points are robust and protected by nonsymmorphic symmetry.

\section*{Acknowledgements}

This work is supported by the National Natural Science Foundation of China (Nos. U2032207, 11974364, U2032208 and 11674367), the National Key Research and Development Program of China (No. 2017YFA0303600), the Natural Science Foundation of Zhejiang, China (No. LZ18A040002), and the Ningbo Science and Technology Bureau (No. 2018B10060). 
The ARPES experiments were performed with the approval of the Proposal Assessing Committee of SiP$\cdot$ME$^2$ platform project (Proposal No. 11227902).



\begin{thebibliography}{99}

\bibitem{Science-YLC-2009}
Y. L. Chen, J. G. Analytis, J.-H. Chu, Z. K. Liu, S.-K. Mo, X. L. Qi, H. J. Zhang, D. H. Lu, X. Dai, Z. Fang, S. C. Zhang, I. R. Fisher, Z. Hussain, and Z.-X. Shen, Experimental realization of a three-dimensional topological insulator, Bi$_2$Te$_3$, Science \textbf{325}, 178 (2009).

\bibitem{Science-YLC-2010}
Y. L. Chen, J.-H. Chu, J. G. Analytis, Z. K. Liu, K. Igarashi, H.-H. Kuo, X. L. Qi, S. K. Mo, R. G. Moore, D. H. Lu, M. Hashimoto, T. Sasagawa, S. C. Zhang, I. R. Fisher, Z. Hussain, and Z. X. Shen, Massive Dirac fermion on the surface of a magnetically doped topological insulator, Science \textbf{329}, 659 (2010).

\bibitem{PRB-2011-AAB}
A. A. Burkov, M. D. Hook, and L. Balents, Topological nodal semimetals, Phys. Rev. B 84, 235126 (2011).

\bibitem{PRL-SMY-2012}
S. M. Young, S. Zaheer, J. C. Y. Teo, C. L. Kane, E. J. Mele, and A. M. Rappe, Dirac semimetal in three dimensions, Phys. Rev. Lett. \textbf{108}, 140405 (2012).

\bibitem{PRB-ZW-2012}
Z. Wang, Y. Sun, X.-Q. Chen, C. Franchini, G. Xu, H. Weng, X. Dai, and Z. Fang, Dirac semimetal and topological phase transitions in ${A}_{3}$Bi ($A=$Na, K, Rb), Phys. Rev. B \textbf{85}, 195320 (2012).

\bibitem{Science-ZKL-2014}
Z. K. Liu, B. Zhou, Y. Zhang, Z. J. Wang, H. M. Weng, D. Prabhakaran, S.-K. Mo, Z. X. Shen, Z. Fang, X. Dai, Z. Hussain, and Y. L. Chen, Discovery of a three-dimensional topological Dirac semimetal, Na$_3$Bi, Science \textbf{343}, 864 (2014).

\bibitem{Science-SYX-2015-Dirac}
S.-Y. Xu, C. Liu, S. K. Kushwaha, R. Sankar, J. W. Krizan, I. Belopolski, M. Neupane, G. Bian, N. Alidoust, T.-R. Chang, H.-T. Jeng, C.-Y. Huang, W.-F. Tsai, H. Lin, P. P. Shibayev, F.-C. Chou, R. J. Cava, and M. Z. Hasan, Observation of Fermi arc surface states in a topological metal, Science \textbf{347}, 294 (2015).

\bibitem{PRB-XW-2011}
X. Wan, A. M. Turner, A. Vishwanath, and S. Y. Savrasov, Topological semimetal and Fermi-arc surface states in the electronic structure of pyrochlore iridates, Phys. Rev. B \textbf{83}, 205101 (2011).

\bibitem{NC-SMH-2015}
S.-M. Huang, S.-Y. Xu, I. Belopolski, C.-C. Lee, G. Chang, B. Wang, N. Alidoust, G. Bian, M. Neupane, C. Zhang, S. Jia, A. Bansil, H. Lin, and M. Z. Hasan, A Weyl Fermion semimetal with surface Fermi arcs in the transition metal monopnictide TaAs class, Nat. Commun. \textbf{6}, 7373 (2015).

\bibitem{PRX-WH-2015}
H. Weng, C. Fang, Z. Fang, B. A. Bernevig, and X. Dai, Weyl Semimetal Phase in Noncentrosymmetric Transition-Metal Monophosphides, Phys. Rev. X \textbf{5}, 011029 (2015).


\bibitem{Science-SYX-2015-Weyl}
S.-Y. Xu, I. Belopolski, N. Alidoust, M. Neupane, G. Bian, C. Zhang, R. Sankar, G. Chang, Z. Yuan, C.-C. Lee, S.-M. Huang, H. Zheng, J. Ma, D. S. Sanchez, B. Wang, A. Bansil, F. Chou, P. P. Shibayev, H. Lin, S. Jia, and M. Z. Hasan, Discovery of a Weyl fermion semimetal and topological Fermi arcs, Science \textbf{349}, 613 (2015).

\bibitem{Science-DW-2018}
D. Wang, L. Kong, P. Fan, H. Chen, S. Zhu, W. Liu, L. Cao, Y. Sun, S. Du, J. Schneeloch, R. Zhong, G. Gu, L. Fu, H. Ding, and H.-J. Gao, Evidence for Majorana bound states in an iron-based superconductor, Science \textbf{362}, 333 (2018).

\bibitem{Science-PZ-2018}
P. Zhang, K. Yaji, T. Hashimoto, Y. Ota, T. Kondo, K. Okazaki, Z. Wang, J. Wen, G. D. Gu, H. Ding, and S. Shin, Observation of topological superconductivity on the surface of an iron-based superconductor, Science \textbf{360}, 182 (2018).

\bibitem{PRB-HW-2016}
H. Weng, C. Fang, Z. Fang, and X. Dai, Topological semimetals with triply degenerate nodal points in $\ensuremath{\theta}$-phase tantalum nitride, Phys. Rev. B \textbf{93}, 241202(R) (2016).

\bibitem{PRX-ZZ-2016}
Z. Zhu, G. W. Winkler, Q. Wu, J. Li, and A. A. Soluyanov, Triple Point Topological Metals, Phys. Rev. X \textbf{6}, 031003 (2016).

\bibitem{NC-WG-2018}
W. Gao, X. Zhu, F. Zheng, M. Wu, J. Zhang, C. Xi, P. Zhang, Y. Zhang, N. Hao, W. Ning, and M. Tian, A possible candidate for triply degenerate point fermions in trigonal layered PtBi$_2$, Nat. Commun. \textbf{9}, 3249 (2018).

\bibitem{NC-2016-GB}
G. Bian, T.-R. Chang, R. Sankar, S.-Y. Xu, H. Zheng, T. Neupert, C.-K. Chiu, S.-M. Huang, G. Chang, I. Belopolski, D. S. Sanchez, M. Neupane, N. Alidoust, C. Liu, B. Wang, C.-C. Lee, H.-T. Jeng, C. Zhang, Z. Yuan, S. Jia, A. Bansil, F. Chou, H. Lin, and M. Z. Hasan, Topological nodal-line fermions in spin-orbit metal PbTaSe$_2$, Nat. Commun. \textbf{7}, 10556 (2016).

\bibitem{PRB-2016-MN}
M. Neupane, I. Belopolski, M. M. Hosen, D. S. Sanchez, R. Sankar, M. Szlawska, S.-Y. Xu, K. Dimitri, N. Dhakal, P. Maldonado, P. M. Oppeneer, D. Kaczorowski, F. Chou, M. Z. Hasan, and T. Durakiewicz, Observation of topological nodal fermion semimetal phase in ZrSiS, Phys. Rev. B \textbf{93}, 201104(R) (2016).

\bibitem{NC-2016-LMS}
L. M. Schoop, M. N. Ali, C. Straßer, A. Topp, A. Varykhalov, D. Marchenko, V. Duppel, S. S. P. Parkin, B. V. Lotsch, and C. R. Ast, Dirac cone protected by non-symmorphic symmetry and three-dimensional Dirac line node in ZrSiS, Nat. Commun. \textbf{7}, 1 (2016).

\bibitem{NP-2018-SP}
S. Pezzini, M. R. van Delft, L. M. Schoop, B. V. Lotsch, A. Carrington, M. I. Katsnelson, N. E. Hussey, and S. Wiedmann, Unconventional mass enhancement around the Dirac nodal loop in ZrSiS, Nat. Phys. \textbf{14}, 2 (2018).

\bibitem{SA-2019-BB}
B.-B. Fu, C.-J. Yi, T.-T. Zhang, M. Caputo, J.-Z. Ma, X. Gao, B. Q. Lv, L.-Y. Kong, Y.-B. Huang, P. Richard, M. Shi, V. N. Strocov, C. Fang, H.-M. Weng, Y.-G. Shi, T. Qian, and H. Ding, Dirac nodal surfaces and nodal lines in ZrSiS, Sci. Adv. \textbf{5}, eaau6459 (2019).

\bibitem{PRL-2020-YKS}
Y. K. Song, G. W. Wang, S. C. Li, W. L. Liu, X. L. Lu, Z. T. Liu, Z. J. Li, J. S. Wen, Z. P. Yin, Z. H. Liu, and D. W. Shen, Photoemission Spectroscopic Evidence for the Dirac Nodal Line in the Monoclinic Semimetal SrAs$_3$, Phys. Rev. Lett. \textbf{124}, 056402 (2020).

\bibitem{NM-2020-TYY}
T. Y. Yang, Q. Wan, D. Y. Yan, Z. Zhu, Z. W. Wang, C. Peng, Y. B. Huang, R. Yu, J. Hu, Z. Q. Mao, S. Li, S. A. Yang, H. Zheng, J.-F. Jia, Y. G. Shi, and N. Xu, Directional massless Dirac fermions in a layered van der Waals material with one-dimensional long-range order, Nat. Mater. \textbf{19}, 1 (2020).

\bibitem{PRB-2016-QFL}
Q.-F. Liang, J. Zhou, R. Yu, Z. Wang, and H. Weng, Node-surface and node-line fermions from nonsymmorphic lattice symmetries, Phys. Rev. B \textbf{93}, 085427 (2016).

\bibitem{PRB-2018-WW}
W. Wu, Y. Liu, S. Li, C. Zhong, Z.-M. Yu, X.-L. Sheng, Y. X. Zhao, and S. A. Yang, Nodal surface semimetals: Theory and material realization, Phys. Rev. B \textbf{97}, 115125 (2018).

\bibitem{PRB-2016-YXZ}
Y. X. Zhao and A. P. Schnyder, Nonsymmorphic symmetry-required band crossings in topological semimetals, Phys. Rev. B \textbf{94}, 195109 (2016).

\bibitem{PRB-2017-BJY}
B.-J. Yang, T. A. Bojesen, T. Morimoto, and A. Furusaki, Topological semimetals protected by off-centered symmetries in nonsymmorphic crystals, Phys. Rev. B \textbf{95}, 075135 (2017).

\bibitem{APX-2018-SYY}
S.-Y. Yang, H. Yang, E. Derunova, S. S. P. Parkin, B. Yan, and M. N. Ali, Symmetry demanded topological nodal-line materials, Advances in Physics: X \textbf{3}, 1414631 (2018).

\bibitem{EJIC-1999-PA}
P. Alemany and E. Canadell, Te...Te Interlayer Interactions, Te Metal Electron Transfer and Electrical Conductivity in the MM'Te5 Phases (M=Nb, M'=Ni, Pd; M=Ta, M'=Ni, Pt), Eur. J. Inorg. Chem. 1701 (1999).

\bibitem{PRB-2020-WHJ}
W.-H. Jiao, X.-M. Xie, Y. Liu, X. Xu, B. Li, C.-Q. Xu, J.-Y. Liu, W. Zhou, Y.-K. Li, H.-Y. Yang, S. Jiang, Y. Luo, Z.-W. Zhu, and G.-H. Cao, Topological Dirac states in a layered telluride TaPdTe$_5$ with quasi-one-dimensional PdTe$_2$ chains, Phys. Rev. B \textbf{102}, 075141 (2020).

\bibitem{JPCL-2020-CX}
C. Xu, Y. Liu, P. Cai, B. Li, W. Jiao, Y. Li, J. Zhang, W. Zhou, B. Qian, X. Jiang, Z. Shi, R. Sankar, J. Zhang, F. Yang, Z. Zhu, P. Biswas, D. Qian, X. Ke, and X. Xu, Anisotropic transport and quantum oscillations in the quasi-one-dimensional TaNiTe$_5$: Evidence for the nontrivial band topology, J. Phys. Chem. Lett. \textbf{11}, 7782 (2020).

\bibitem{PRB-2021-WHJ}
W.-H. Jiao, S. Xiao, B. Li, C. Xu, X.-M. Xie, H.-Q. Qiu, X. Xu, Y. Liu, S.-J. Song, W. Zhou, H.-F. Zhai, X. Ke, S. He, and G.-H. Cao, Anisotropic transport and de Haas--van Alphen oscillations in quasi-one-dimensional TaPtTe$_5$, Phys. Rev. B \textbf{103}, 125150 (2021).

\bibitem{PRB-2021-ZC}
Z. Chen, M. Wu, Y. Zhang, J. Zhang, Y. Nie, Y. Qin, Y. Han, C. Xi, S. Ma, X. Kan, J. Zhou, X. Yang, X. Zhu, W. Ning, and M. Tian, Three-dimensional topological semimetal phase in layered TaNiTe$_5$ probed by quantum oscillations, Phys. Rev. B \textbf{103}, 035105 (2021).

\bibitem{PRB-2021-ZH}
Z. Hao, W. Chen, Y. Wang, J. Li, X.-M. Ma, Y.-J. Hao, R. Lu, Z. Shen, Z. Jiang, W. Liu, Q. Jiang, Y. Yang, X. Lei, L. Wang, Y. Fu, L. Zhou, L. Huang, Z. Liu, M. Ye, D. Shen, J. Mei, H. He, C. Liu, K. Deng, C. Liu, Q. Liu, and C. Chen, Multiple dirac nodal lines in an in-plane anisotropic semimetal ${\mathrm{TaNiTe}}_{5}$, Phys. Rev. B \textbf{104}, 115158 (2021).


\bibitem{SSRF-1}
Z. P. Sun, Z. H. Liu, Z. T. Liu, W. L. Liu, F. Y. Zhang, D. W. Shen, M. Ye, and S. Qiao, Performance of the BL03U beamline at SSRF, J. Synchrotron Rad. \textbf{27}, 1388 (2020).

\bibitem{SSRF-2}
Y.-C. Yang, Z.-T. Liu, J.-S. Liu, Z.-H. Liu, W.-L. Liu, X.-L. Lu, H.-P. Mei, A. Li, M. Ye, S. Qiao, and D.-W. Shen, High-resolution ARPES endstation for in situ electronic structure investigations at SSRF, Nucl. Sci. Technol. \textbf{32}, 31 (2021).

\bibitem{JPCM-QE-2009}
P. Giannozzi, S. Baroni, N. Bonini, M. Calandra, R. Car, C. Cavazzoni, D. Ceresoli, G. L. Chiarotti, M. Cococcioni, I. Dabo, A. D. Corso, S. de Gironcoli, S. Fabris, G. Fratesi, R. Gebauer, U. Gerstmann, C. Gougoussis, A. Kokalj, M. Lazzeri, L. Martin-Samos, N. Marzari, F. Mauri, R. Mazzarello, S. Paolini, A. Pasquarello, L. Paulatto, C. Sbraccia, S. Scandolo, G. Sclauzero, A. P. Seitsonen, A. Smogunov, P. Umari, and R. M. Wentzcovitch, QUANTUM ESPRESSO: a modular and open-source software project for quantum simulations of materials, J. Phys.: Condens. Matter. \textbf{21}, 395502 (2009).

\bibitem{PRB-PAW-1994}
P. E. Blöchl, Projector augmented-wave method, Phys. Rev. B \textbf{50}, 17953 (1994).

\bibitem{PRL-PBE-1996}
J. P. Perdew, K. Burke, and M. Ernzerhof, Generalized Gradient Approximation Made Simple, Phys. Rev. Lett. \textbf{77}, 3865 (1996).

\bibitem{JPCM-wannier90-2020}
G. Pizzi, V. Vitale, R. Arita, S. Blügel, F. Freimuth, G. Géranton, M. Gibertini, D. Gresch, C. Johnson, T. Koretsune, J. Ibañez-Azpiroz, H. Lee, J.-M. Lihm, D. Marchand, A. Marrazzo, Y. Mokrousov, J. I. Mustafa, Y. Nohara, Y. Nomura, L. Paulatto, S. Poncé, T. Ponweiser, J. Qiao, F. Thöle, S. S. Tsirkin, M. Wierzbowska, N. Marzari, D. Vanderbilt, I. Souza, A. A. Mostofi, and J. R. Yates, Wannier90 as a community code: new features and applications, J. Phys.: Condens. Matter. \textbf{32}, 165902 (2020).

\bibitem{CPC-wt-2018}
Q. Wu, S. Zhang, H.-F. Song, M. Troyer, and A. A. Soluyanov, WannierTools: An open-source software package for novel topological materials, Comput. Phys. Commun. \textbf{224}, 405 (2018).

\bibitem{supplementary}
See Supplemental Material at ... for more details of theoretical symmetry analysis, photon-energy-dependent ARPES measurements and DFT calculations.

\bibitem{PRB-2021-Xiao}
S. Xiao, Y. Li, Y. Li, X. Yang, S. Zhang, W. Liu, X. Wu, B. Li, M. Arita, K. Shimada, Y. Shi, and S. He, Direct evidence of electron-hole compensation for extreme magnetoresistance in topologically trivial YBi, Phys. Rev. B \textbf{103}, 115119 (2021).

\bibitem{PRB-2019-Wu}
Z. Wu, F. Wu, P. Li, C. Guo, Y. Liu, Z. Sun, C.-M. Cheng, T.-C. Chiang, C. Cao, H. Yuan, Y. Liu, Probing the origin of extreme magnetoresistance in Pr/Sm mono-antimonides/bismuthides, Phys. Rev. B \textbf{99}, 035158 (2019).

\bibitem{PRB-2018-Wang}
Y.-Y. Wang, H. Zhang, X.-Q. Lu, L.-L. Sun, S. Xu, Z.-Y. Lu, K. Liu, S. Zhou, T.-L. Xia, Extremely large magnetoresistance and electronic structure of TmSb, Phys. Rev. B \textbf{97}, 085137 (2018).

\bibitem{PRB-2016-Niu}
X. H. Niu, D. F. Xu, Y. H. Bai, Q. Song, X. P. Shen, B. P. Xie, Z. Sun, Y. B. Huang, D. C. Peets, and D. L. Feng, Presence of exotic electronic surface states in LaBi and LaSb, Phys. Rev. B \textbf{94}, 165163 (2016).

\bibitem{PRM-2018-Jiang}
J. Jiang, N. B. M. Schröter, S.-C. Wu, N. Kumar, C. Shekhar, H. Peng, X. Xu, C. Chen, H. F. Yang, C.-C. Hwang, S.-K. Mo, C. Felser, B. H. Yan, Z. K. Liu, L. X. Yang, and Y. L. Chen, Observation of topological surface states and strong electron/hole imbalance in extreme magnetoresistance compound LaBi, Phys. Rev. Materials \textbf{2}, 024201 (2018).

\bibitem{PRB-2017-Oinuma}
H. Oinuma, S. Souma, D. Takane, T. Nakamura, K. Nakayama, T. Mitsuhashi, K. Horiba, H. Kumigashira, M. Yoshida, A. Ochiai, T. Takahashi, and T. Sato, Three-dimensional band structure of LaSb and CeSb: Absence of band inversion, Phys. Rev. B \textbf{96}, 041120(R) (2017).

\end{thebibliography}
\end{document}